# Generation of Accelerating Waves in Smith-Purcell Radiation

LIQIAO JING,[1,2,4] DASHUANG LIAO, [1,2,4] JIE TAO, [1,2] HONGSHENG CHEN, [1,2,3,*] AND ZUOJIA WANG[1,2,*]

[1]*Interdisciplinary Center for Quantum Information, State Key Laboratory of Modern Optical Instrumentation, ZJU-Hangzhou Global Scientific and Technological Innovation Center, Zhejiang University, Hangzhou 310027, China.*
[2]*International Joint Innovation Center, Key Lab. of Advanced Micro/Nano Electronic Devices & Smart Systems of Zhejiang, The Electromagnetics Academy at Zhejiang University, Zhejiang University, Haining 314400, China*
[3]*Jinhua Institute of Zhejiang University, Zhejiang University, Jinhua 321099, China*
[4]*These authors contributed equally to this work.*
*\*Corresponding author: zuojiawang@zju.edu.cn; hansomchen@zju.edu.cn*

**Metasurface has recently emerged as a powerful platform to engineer wave packets of free electron radiation at the mesoscale. Here, we propose that accelerating waves can be generated when moving electrons interact with an array of bianisotropic meta-atoms. By changing the intrinsic coupling strength, we show full amplitude coverage and 0-to-π phase switching of Smith-Purcell radiation from bianistropic meta-atoms. This unusual property leads to the creation of Airy beams that shifts along a parabolic trajectory during propagation. Experimental implementation displays that evanescent fields bounded at slotted waveguides can be coupled into accelerating waves via Smith-Purcell radiation from a designer bianisotropic metasuface. Our results offer an alternative route towards free electron lasers with diffraction-free, self-accelerating, and self-healing beam properties. © 2020 Optica Publishing Group**

## 1. INTRODUCTION

Accelerated electrons can emit electromagnetic waves when they pass through a medium, which is known as free-electron radiation [1]. Typical classes of free-electron radiation include Cherenkov radiation [2-10] for electrons traveling faster than the phase velocity of light, Smith-Purcell radiation [11-18] for low-speed electrons on periodic structures, and transition radiation [19-22] for electrons passing through inhomogeneous media. Free electron radiation plays a pivotal role in many practical applications including particle detectors [23, 24], particle accelerators [25-27], free-electron lasers [28-30], bioimaging and security detection [31]. One trend that has emerged recently is the use of two-dimensional optical interfaces, such as metasurfaces [32, 33] and 2D materials [34-36], to control at will the light-matter interaction of free-electron radiation. C-aperture Babinet metasurfaces [37, 38] have been proposed to control the polarization direction of Smith-Purcell radiation through induced cross-coupled electric and magnetic dipoles. Graphene metasurfaces have been designed to manipulate the intensity and polarization states of the radiated waves by turning the structure and Fermi level [39]. A moving charge interacting with a hyperbolic material can lead to Cherenkov radiation without any predefined value of the kinetic energy of the electron [8]. Tunable free-electron X-ray radiation from van der Waals materials has been demonstrated in practice [40]. Resonance transition radiation in photonic crystals has been proposed to achieve highly sensitive particle detectors with miniaturized volume [6].

Airy beams, as self-accelerating optical waves, were first predicted in the context of quantum mechanics and then introduced to the optics using analogy between the Schrodinger equation and the paraxial Helmholtz equation [41, 42]. The intensity of an Airy beam follows a curved parabolic trajectory, providing remarkable diffraction-free, self-accelerating, and self-healing properties [43, 44]. The concept of accelerating wave packets inspires variant potential applications such as forming optical bullet [45, 46], optical micro-manipulation [47], plasma channel generating [48], laser micromachining [49], signal processing [50] and so on. Airy beams are usually realized by phase masks formed by spatial light modulators, or diffractive optical elements and an extra Fourier-transform lens [51-53]. Benefiting from the flexibility in amplitude and phase modulation, metasurfaces have been adopted to generate non-diffraction and accelerating optical waves including Airy beams [54], Bessel beams [55-58], and so on. However, the interaction between accelerating optical waves and accelerated electrons at optical interfaces is still less studied.

In this paper, we propose a novel method to generate and steer accelerating waves from moving electrons. The amplitude and phase profiles of accelerating waves in Smith Purcell radiation can be mapped to the intrinsic coupling strength in a three-layered meta-atom, which can be tailored by changing the rotation angle of the I-shape resonator in the middle layer (Fig. 1). The bianisotropy of the designed meta-atom contributes to full amplitude coverage and 0-to-π phase switching, as well as providing polarization conversion. Moreover, the meta-atom shows dispersionless phase responses and high-efficiency radiation, thus conductive to implementing accelerating wave in Smith Purcell radiation over a broadband range. Numerical and experimental results show that the evanescent fields bounded at moving electrons and slotted waveguides can be radiated into accelerating waves through the designer bianisotropic metasurface.

## 2. RESULTS

Swift electrons with charge $q$ are assumed to move at a plane below the bianisotropic metasurface, with the electron velocity $\vec{v}_0 = \vec{z}v_0 = \vec{z}\beta c$, where $c$ is the speed of light in free space and $\beta$ is the velocity normalized by the speed of light ($\beta = v_0/c$). The induced current density by moving electrons in time domain is $\vec{J}(\vec{r},t) = \vec{z}qv\delta(y-y_0)\delta(z-vt)$. After Fourier transformation, the induced current density in frequency domain can be expressed as $\vec{J}(\vec{r},\omega) = \frac{1}{2\pi}\int \vec{J}(\vec{r},t)e^{i\omega t}dt = \vec{z}\frac{qv}{2\pi}\delta(y-y_0)e^{ik_z z}$, where $k_z = \omega/v$. Such a form of current density in frequency domain also exists in phased dipole arrays or slotted waveguides, as demonstrated in Ref. [3].

The key point to generate Airy beams in Smith Purcell radiation is to design meta-atoms with a free combination of arbitrary amplitude and binary phase responses. This can be realized by using a three-layered metasurface as illustrated in Fig. 2(a). Both the top and bottom layers are one-dimensional (1D) metallic gratings. These two sets of gratings are orthogonal to each other, and the neighboring metal strips are separated by an air gap of 9 mm. The middle layer consists of an array of symmetric split-ring resonators connected by a cut wire, where $α$ represents the azimuthal rotation angle about the y axis. Three metallic layers are separated by two F4B slabs, with the thickness of 2 mm and a relative permittivity of $\varepsilon_r = 2.65 + 0.01i$ for each one. Copper is adopted as the metal with a conductivity $\sigma = 5.8\times 10^7$ S/m and a thickness of 0.035 mm. The meta-atom has a periodicity p = 20 mm. In such meta-atoms, continuous amplitude and binary phase modulation can be simultaneously achieved by varying the value of $α$.

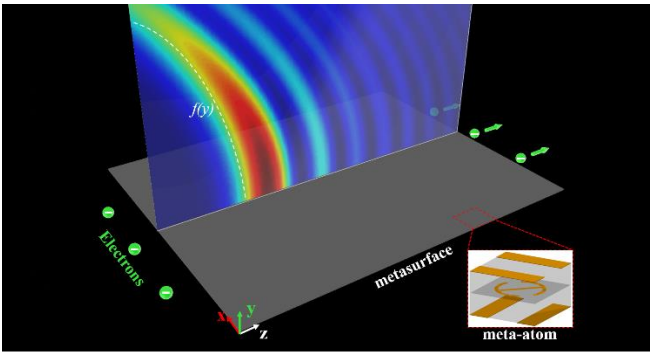

Fig. 1. Schematic of Airy beams in Smith Purcell radiation from amplitude and phase controllable bianisotropic metasurface. Here and below, a uniform sheet of electrons moves with a velocity $\vec{v} = v_0\vec{z}$ along a trajectory parallel to the metasurface plane, where $v_0 = 0.36c$.

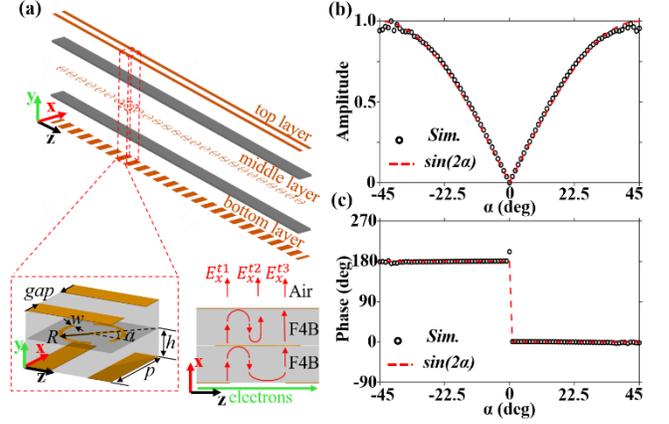

Fig. 2. Analysis and performance of the proposed meta-atoms. (a) Schematic of full amplitude coverage and 0-to-π phase switching bianisotropic metasurface. In each meta-atom of the middle layer, the symmetric split ring has an open angle and the central line of opening (black dash line) is azimuthally rotated by an angle α with respect to the z axis. (b) (c) Theoretical (red dash line) and simulated (black dots) results of amplitude (b) and phase (c) of the Smith Purcell radiation at 5.4 GHz. Caption text with descriptions of (a), (b), and (c).

We first study the Smith-Purcell radiation effect from the middle resonator layer. When swift electrons with the charge $q$ move with an arbitrary direction in the $xoz$ plane, the radiated electric intensity exhibits a Jones matrix:

$$E_{ex}(\alpha) = S(-\alpha) \cdot E_{in} \cdot S(\alpha) \quad (1)$$

in which $E_{in} = \begin{pmatrix} E_{zz} & 0 \\ 0 & E_{xx} \end{pmatrix}$ represents the radiated electric matrix when $α = 0$, $E_{xx}$ and $E_{zz}$ are the x-polarized and z-polarized radiated fields from swift electrons moving along the x and z directions, respectively. $S(\alpha) = \begin{pmatrix} \cos\alpha & \sin\alpha \\ -\sin\alpha & \cos\alpha \end{pmatrix}$ represents the rotation matrix when the middle-layer resonator rotates. As a result, the cross polarized radiation of the rotated meta-atom is given by:

$$E_{xz} = \frac{(E_{zz} - E_{xx})}{2}\sin(2\alpha) \quad (2)$$

Eq. (2) implies that the intensity of cross-polarized field is always positively correlated to the sine of the orientation angle. This provides a possible means to achieve controllable amplitude responses by varying the rotation angle of the meta-atom. However, a single-layer metasurface with barely electric resonances can only achieve 25% polarization conversion efficiency. The efficiency can be improved by introducing additional two layers, based on the principle of Fabry-Perot resonance. Eq. (2) also implies that the maximum intensity can be achieved when $α$ is set as 45° when the electron moves along z axis. Due to the polarization effect of the top and bottom wire gratings, the co-polarization components will be suppressed and we then obtain

$$E_{xz} \propto E_{zz} \sin(2\alpha) \quad (3)$$

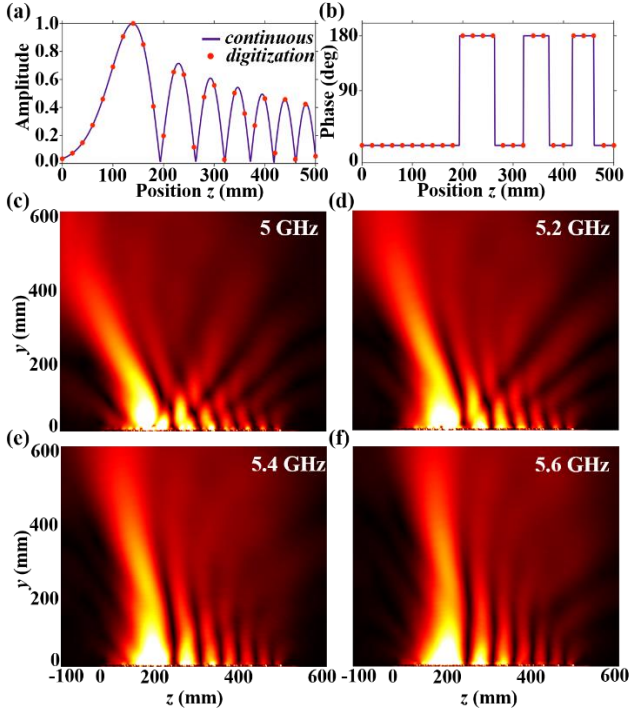

Fig. 3. (a) Amplitude and (b) phase profiles for 1D Airy beam in Smith Purcell radiation along the z-direction. (c)-(f) Simulated electric distribution (Ex) on the *yoz* plane at (c) 5.0 GHz, (d) 5.2 GHz, (e) 5.4 GHz and (f) 5.6 GHz, respectively.

Therefore, continuous amplitude modulation and 0-to-π phase switching can be simultaneously achieved by changing the rotation angle α.

Under the guidance of the above principle, full-wave simulations have been performed in commercial software, COMSOL MULTIPHYSICS. In the simulation, periodic boundary is applied in the *x*- and *z*-directions and the swift electrons are set as $v_0=0.36c$ moving along the +z direction and the current density is I0 = 1 A/m. Under these conditions, Smith Purcell radiation at 5.4 GHz propagates along the *y* direction where $k_z = 0$. As shown in Fig. 2a, the maximum radiation intensity occurs when the parameters are optimized as: radius of the circle is R = 7 mm, the width of the connecting rod is fixed as 1 mm, the open cut angle is 35°. The simulated amplitude and phase responses of the radiated wave at 5.4 GHz versus the orientation angle α are plotted in Figs. 2(b) and 2(c). The radiation phases are all normalized to the case of α = -45°. These simulation results show full amplitude coverage from 0 to 1, and phase switching between 0 and π, in good agreement with the theoretical prediction from Eq. (3). This exactly satisfies the requirements on wavefront modulation to generate Airy beams in Smith Purcell radiation, which will be discussed later.

According to the non-diffraction property of the Airy beams, the amplitude and phase distributions across the metasurface profile (at the plane of y = 0) are required by:

$$\phi(z,0) = Ai(z/\omega) \cdot \exp(a \cdot z/\omega) \quad (4)$$

where *Ai* is an Airy function, *a* is the truncation factor to guarantee infinite Airy tail, *w* represents the scaling length. The parameters in Eq. (4) are fixed as a = 0.04 and w = 0.04. To implement the Airy beams in Smith Purcell radiation based on bianisotropic metasurface, the amplitude and phase profiles on the bianisotropic metasurface should be discretized by a lattice constant p = 20 mm. The sample range is designed from 0 mm to 500 mm and the sample number is 26. For generation of 1D Airy beam, the meta-atoms along the *x* direction are identical. The designed amplitude and phase distributions along the z direction are shown in Fig. 3(a) and 3(b), respectively. The violet line shows the continuous distribution calculated by Eq. (4) and the red dots correspond to the discretized ones with a period of 20 mm.

Figs. 3(c-f) show the simulated field distributions of the emitted light from the metasurface. It is notable that the ballistic trajectory propagation direction depends on both the meta-atom distribution and the operating frequency. In order to clarify this influence, we set the periodicity $p=-m\lambda/c$ at 5.4 GHz, where the order of m = -1 is radiated according to the generalized Smith Purcell formula [18]. The trajectory curvature will be larger when the wavelength increases. Moreover, the polarization of the emitted light can be modulated by the bianisotropic metasurface, in consistent with the above theoretical analysis. Interesting, the designer metasurface radiates cross-polarized waves that cannot be achieved in conventional gratings. The underlying mechanism is spatially extended interaction between moving electrons and the grating along the z direction can only induce a bunch of electric dipoles oscillating along the same direction.

Furthermore, the proposed meta-atoms can be readily extended to generate 2D Airy beams in Smith Purcell radiation by satisfying the following equation:

$$\phi = Ai(z/\omega_1) \cdot Ai(x/\omega_2) \cdot \exp(a \cdot z/\omega_1) \cdot \exp(a \cdot x/\omega_2) \quad (5)$$

The parameters of Eq. (5) are selected as $a_1 = a_2 = 0.04$ and $w_1 = w_2 = 0.04$. Here, 26×26 unit cells cover from 0 mm to 500 mm in the x and the z directions. The designed amplitude and phase distributions are shown in Fig. 4(a) and 4(b), respectively. The numerical simulated results for 2D Airy beam in Smith Purcell radiation are presented in Fig. 4(c), for different planes of y = 100 mm, 200 mm, 300 mm, 400 mm and 500 mm. As the propagation distance increases, the main lobe of the 2D Airy Smith Purcell radiation moves along the diagonal direction of the plane, showing the self-acceleration and non-diffraction characteristics.

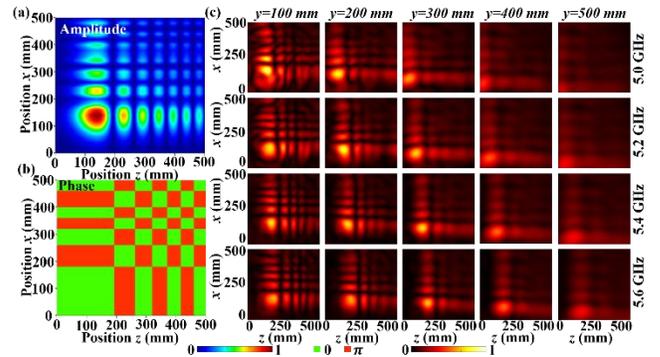

Fig. 4. (a) Amplitude and (b) phase profiles for 2D Airy beam in Smith Purcell radiation along x- and z- directions. (c) Simulated electric distribution (Ex) on the xoz plane at 5.0 GHz, 5.2 GHz, 5.4 GHz and 5.6 GHz at different planes y = 100 mm, 200 mm, 300 mm, 400 mm and 500 mm, respectively.

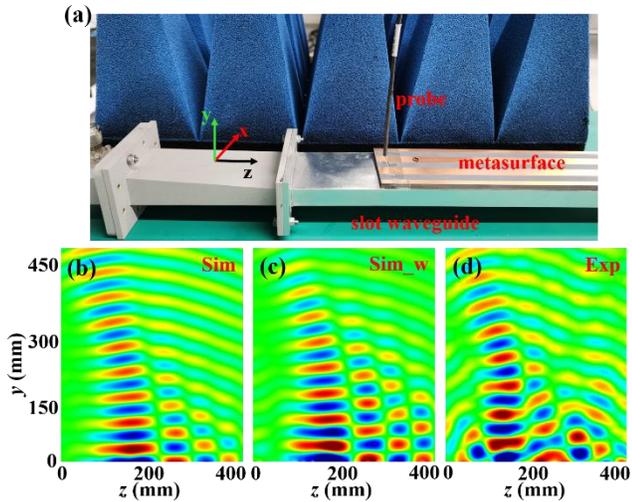

Fig. 5. Experimental observation of Airy beams in Smith Purcell radiation (a) Experimental measurement setup. (b-c) Numerically electric distribution (Ex) from realistic moving electrons (b) and phased dipole array in slotted waveguide (c). (d) Experimental measurement result of electric distribution (Ex) at 5.4 GHz.

Fig. 5 shows the experimental observation of the Airy beam in Smith Purcell radiation. The experimental setup is shown in Fig. 5(a). In the microwave experiment, the slotted waveguide is placed 0.5 mm beneath the bottom layer of the bianisotropic metasurface. Evanescent fields bounded at the slotted waveguide can be coupled into accelerating waves via Smith-Purcell radiation. The amplitude and phase distributions of emitted light have been simultaneously measured by using the near-field scanning platform in an anechoic chamber. Due to the radiated power from electrons at the microwave frequency is generally small. To facilitate the experimental observation of the Airy beams in Smith Purcell radiation, swift electrons can be effectively mimicked by the phased dipole array and due to the inside dielectric with permittivity 9.2 and then the phase velocity of $v_{phase} = 0.36c$, which is used to effectively model the moving electrons with the desired velocity (i.e., $v = v$phase) at the studied frequency of 5.4 GHz. We have compared the measured radiation fields (Fig. 5(d)) with those numerically studied from realistic moving electrons (Fig. 5(b)) and phased dipole array in slotted waveguide (Fig. 5(c)). As expected, all results demonstrate the shifting of the main lobe along a parabolic trajectory during propagation. Perhaps one of the most exciting applications of accelerating wave in Smith Purcell radiation is particle manipulation which not only fueled further research interest in accelerating beams, but also added a new tool in the arena of optical trapping and manipulation, where light-induced forces play a crucial role.

## 2. DISCUSSION

In conclusion, we propose that accelerating waves can be generated when moving electrons interact with an array of bianisotropic meta-atoms which can module arbitrary continuous amplitude and phase. Free combination with binary phase responses can be easily achieved by tailoring only intrinsic coupling strength. Moreover, we have experimentally observed the polarization shaping of the free-electron radiation to generate Airy beams in Smith Purcell radiation using bianisotropic metasurfaces. The exploited technique for the flexible shaping of free-electron radiation should be applicable in other frequencies ranging from terahertz to ultraviolet. Therefore, our work further indicates that the bianisotropic metasurface can provide a promising versatile platform for generating remarkable diffraction-free, self-accelerating, and self-healing free-electron radiation and may boost the development of future on-chip nanophotonic circuits and devices, such as on-chip free-electron light sources as well as miniaturized particle accelerators, and novel sensing and diagnostic devices.

**Funding.** National Natural Science Foundation of China (62171407, 11961141010, and 61975176).

**Disclosures**. The authors declare no conflicts of interest

**Data availability**. Data underlying the results presented in this paper are not publicly available at this time but may be obtained from the authors upon reasonable request.